# Simulation of stress-impedance effects in low magnetostrictive films

B. Kaviraj[*] and S. K. Ghatak

Department of Physics & Meteorology, Indian Institute of Technology, Kharagpur 721302, India

## Abstract

A theoretical study of stress-impedance effect based on the solution of Landau-Lifsitz-Gilbert equation has been carried out. The results show that stress impedance effects depend largely on several extrinsic (external bias field, external frequency) and intrinsic (orientation and magnitude of uniaxial anisotropy, damping) parameters.

Keywords: Stress-impedance, Magneto-impedance, Magnetostriction, Landau-Lifsitz-Gilbert equation.

---
[*] Corresponding author
Email: bhaskar@phy.iitkgp.ernet.in
(B. Kaviraj)

# 1. Introduction

Low magnetostrictive amorphous alloys have attractive magnetic properties in the high frequency range as high permeability, high saturation magnetic induction and low magnetic losses that make these materials suitable for observing the magneto-impedance (MI) and stress-impedance (SI) effects [1-5]. In recent years, the stress-impedance effects based on the magnetoelastic films showing the inverse magnetoelastic effect have been extensively studied because of scientific interests and industrial applications. As is well-known, the magneto-impedance (MI) effect is the variation of a.c impedance of a ferromagnetic conductor in presence of an external field. The origin of MI is related to the behaviour of the permeability, μ, at high frequency with the applied magnetic field [6, 7] through the skin penetration depth, $\delta = \dfrac{c}{\sqrt{2\pi\sigma_c \omega\mu}}$ where $\sigma_c$ and μ are the conductivity and permeability of the material, ω is the angular frequency and c is the speed of light. The permeability of a ferromagnetic conductor depends upon several factors, such as the external bias field, excitation frequency and current amplitude, external stress, magnetic anisotropy, heat treatment, etc. A tensile stress applied to a nearly zero magnetostrictive ribbon changes its saturation magnetostriction coefficient, due to the magnetoelastic interaction that is related to the strain dependence of the magnetic anisotropy energy [8]. This stress dependence of saturation magnetostriction coefficient together with the applied stress affects the effective anisotropy field and anisotropy of the ribbons. Hence a stress-impedance (SI) effect appears in the samples when a high frequency current generates the skin effect [9, 10].

Most of the studies on magnetoelastic materials are about experiments and there are relatively fewer theoretical studies as we know. In this paper, a skin effect based explanation to the SI phenomenon will be presented. Based on the skin effect explanation and the solution of Landau-Lifsitz-Gilbert equation, the stress dependence of impedance of a magnetoelastic film upon external magnetic field, external frequency, orientation of easy axis, magnitude of uniaxial anisotropy and damping parameters have been simulated.

## 2. MODEL AND FORMULATION

It is well known that the AC current is not homogenous over the cross-section of the magnetic conductor due to the screening of e.m field. The screening is governed by the Maxwell equations along with the magnetization dynamics. In soft ferromagnetic materials, *M* becomes a non-linear function of *H* and this leads to nonlinear coupled equations for *M* and *H*. Assuming the linear magnetic response of the material, the above coupled equation can be solved and inhomogeneous distribution of the current is then characterized by the skin depth

$$\delta = \sqrt{\frac{1}{\sigma_c f \mu_{eff}}} \qquad (1)$$

where 'f' is the frequency of the AC current, $\sigma_c$ is the electrical conductivity and $\mu_{eff}$ is the effective permeability of the magnetic film. In magnetic films, the permeability $\mu_{eff}$ depends upon the frequency f, the amplitude of the bias magnetic field, the applied stress, etc. The impedance 'Z' of a long magnetic ribbon of thickness '2d' which has been excited by a current carrying signal coil of inductance '$L_0$' wounded across the ribbon is given by:

$$Z = -i\omega L_0 \mu \qquad (2)$$

where $\mu = \mu_{eff} \dfrac{tanh(kd)}{kd}$ \qquad (3)

with $k = (1+i)/\delta$ \qquad (4)

and $L_0 = \dfrac{n^2 A_c}{2l} \mu_0$ is the inductance of the empty coil and $\mu_{eff}$ is the effective permeability.

The permeability $\mu_{eff}$ of the ferromagnetic material is a complex quantity due to magnetic relaxation and alters the impedance of the sample in a non-linear way in presence of external stress or magnetic field. As the permeability is a measure of magnetic response, it is necessary to consider the magnetization dynamics in presence of small excitation field and external parameters and then to estimate the permeability and its variation with the parameters. The magnetization dynamics of ferromagnetic material in macroscopic scale is customarily described by the Landau-Lifsitz-Gilbert equation:

$$\dot{\vec{M}} = \gamma\left(\vec{M} \times \vec{H}_{eff}\right) - \dfrac{\alpha}{M_s}\left(\vec{M} \times \dot{\vec{M}}\right) - \dfrac{1}{\tau}\left(\vec{M} - \vec{M}_0\right) \qquad (5)$$

Here $M$ is the magnetization, $\gamma$ is the gyromagnetic ratio, $M_s$ the saturation magnetization, $H_{eff}$ is the effective field and $M_0$ the equilibrium magnetization. The first term in right hand side of equation (5) is torque acting on $M$ due to $H_{eff}$, the second term is the Gilbert damping term with damping coefficient $\alpha$. The last term is referred as the Bloch

Bloembergen damping with relaxation time τ. This does not preserve the magnitude of macroscopic magnetization, as is required for an ideal ferromagnet, and is used to describe the relaxation processes in materials with imperfect ferromagnetic order (such as amorphous and nanocrystalline alloys or crystals with some structural defects). It has been proved that the choice of the particular damping term substantially influences the imaginary part of effective permeability and consequently the magnitude of magneto-impedance effect [11]. The effective field $H_{eff}$ can be written as:

$$\vec{H}_{eff} = \vec{H} + \vec{H}_a + \vec{H}_\sigma \qquad (6)$$

where the exchange coupling field and the demagnetizing field have been neglected to simplify the computation. The corresponding fields are defined as follows: **H** is the sum of applied d.c bias field and exciting a.c field. The uniaxial anisotropy field is

$$\vec{H}_a = \frac{2K_u}{\mu_0 M_s^2}\vec{e}_a\left(\vec{e}_a \bullet \vec{M}\right) = \frac{H_k}{M_s}\vec{e}_a\left(\vec{e}_a \bullet \vec{M}\right) \qquad (7)$$

where $e_a$ is the unit vector along the easy axis. $K_u$ is the uniaxial anisotropy constant and

$$H_k = \frac{2K_u}{\mu_0 M_s}$$

The applied stress effective field is

$$\vec{H}_\sigma = \frac{H_{\sigma 1}}{M_s}\vec{e}_\sigma\left(\vec{e}_\sigma . \vec{M}\right) \qquad (8)$$

where $e_\sigma$ is the unit vector along the applied stress direction and

$$H_{\sigma 1} = \frac{3\lambda\sigma}{\mu_0 M_s} \qquad (9)$$

where λ is the magnetostriction coefficient. In presence of low amplitude driving current **I**, the excitation a.c field **h** is much smaller than the other magnetic fields. Therefore, the induced magnetization **m** is small and the deviation of **M** from its equilibrium orientation **M₀** is also small. So one can assume **H**$_{eff}$ = **H**$_{eff0}$ + **h**$_{eff}$ and **M** = **M₀** +**m** and the a.c component of the vectors varies as

**h, h**$_{eff}$**, m** α $e^{i\omega t}$ \qquad (10)

where ω = 2πf is the circular frequency of the a.c current.

From Eqs. (6), (7), (8) and (10) we get

$$\vec{H}_{eff0} = \vec{H} + \frac{H_k}{M_s}\vec{e}_a\left(\vec{e}_a \cdot \vec{M}_0\right) + \frac{H_{\sigma 1}}{M_s}\vec{e}_\sigma\left(\vec{e}_\sigma \cdot \vec{M}_0\right) \qquad (11)$$

$$\vec{h}_{eff} = \vec{h} + \frac{H_k}{M_s}\vec{e}_a\left(\vec{e}_a \cdot \vec{m}\right) + \frac{H_{\sigma 1}}{M_s}\vec{e}_\sigma\left(\vec{e}_\sigma \cdot \vec{m}\right)$$

$$= \vec{h} + \vec{h}_a \qquad (12)$$

Substituting Eqs.(10)-(12) into (5) and then rewriting (5)

$$\frac{i\omega^*}{\gamma}\vec{m} + \left(\frac{i\alpha\omega}{\gamma}\frac{\vec{M}_0}{\vec{M}_s} + \vec{H}_{eff0}\right) \times \vec{m} = \left(\vec{M}_0 \times \vec{h}_{eff}\right) \qquad (13)$$

Here $\omega^* = \omega - i/\tau$, where $\tau$ is related to the relaxation frequency, $\omega_0$, by $\omega_0 = 1/\tau$.

The magnetic ribbon is modeled as a system with a uniform uniaxial in-plane magnetic anisotropy oriented at an angle $\theta_0$ with respect to the z-axis (Fig.1).

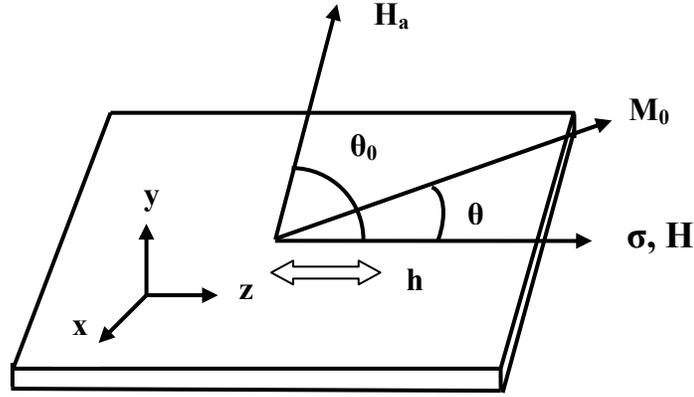

Fig. 1. Schematic view of magnetoelastic film showing different directions of anisotropy fields and stress.

The stress $\sigma$ or d.c field (**H**) is assumed to be applied along z-axis and the magnetization **$M_0$**, in absence of a.c. field, lies in the film plane at an angle $\theta$ with respect to the z-axis. The a.c components of magnetization can be computed from equation (13) and the effective permeability can be obtained by

$$\mu_{eff} = \left(\frac{m_z}{h_z} + 1\right) \qquad (14)$$

The impedance of the magnetoelastic film can then be obtained from the Eqs. (1), (2) and (14). For solution of equation (13), it is necessary to determine the components $M(\theta)$. In the absence of a.c magnetic field, the equilibrium magnetization, as follows from equation (5), is determined by:

$$\left(\vec{M}_0 \times \vec{H}_{eff}\right) = 0 \qquad (15)$$

With this equation, the equilibrium angle θ could be obtained when **H**, **H**$_a$ and **H**$_\sigma$ are given. The equilibrium angle θ is first obtained for different stresses from equation (15) and these values are used to calculate μ$_{eff}$ (equation 14) and hence the impedance Z as functions of frequency, stress and magnetic field is determined. Considering the empirical relation between bulk magnetostriction and magnetization [12], we write

$$\lambda = \gamma_1 M^2 \qquad (16)$$

where γ$_1$ depends upon the stress. The stress dependence of the magnetostriction curve $\lambda(M,\sigma)$ can be described in terms of γ$_1$ and can be expressed in Taylor series in powers of σ [12] as:

$$\gamma_1(\sigma) = \gamma_1(0) + \sum_{n=1}^{\infty} \frac{\sigma^n}{n!} \gamma_1^n(0) \qquad (17)$$

where $\gamma_1^n(0)$ is the nth derivative of γ$_1$ with respect to stress at σ = 0.

## 3. NUMERICAL RESULTS AND DISCUSSION

Based on the above theoretical model and using typical values of parameters $\theta_0 = 50^0$, $H_a = 200$ A/m, $\alpha = 20$, $M_s = 6.5 \times 10^5$ A/m and $\omega_0 = 10^6$ rad/s, we first present the frequency response of real and imaginary components of impedance as functions of different applied stresses. This is shown in Fig.2. The values of $\gamma_1(0) = -7 \times 10^{-18} A^{-2} m^2$ and $\gamma_1'(0) = 1 \times 10^{-29} A^{-2} m^2 Pa^{-1}$ were used in all the cases. These values are typical for soft ferromagnetic materials [12]. The value of $H_a$ chosen was close to the anisotropy fields of FeCoSiB alloys [13].

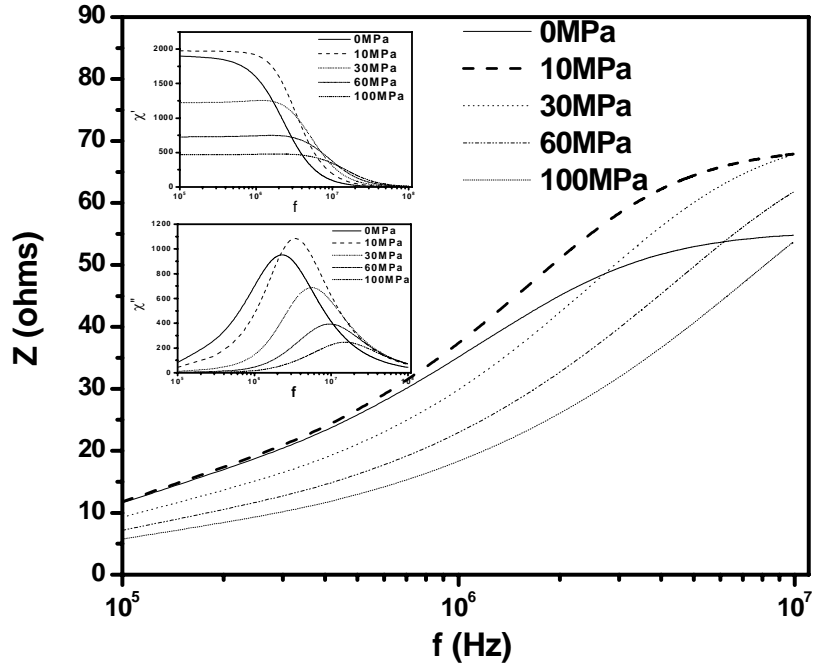

Fig.2. Frequency dependence of Z as functions of applied stresses. The values of parameters taken are $\theta_0 = 50^0$, $H_a = 200$A/m, $\alpha=20$, $M_s = 6.5 \times 10^5$ A/m, and $\omega_0 = 10^6$ rad/s. Inset shows the frequency response of susceptibility at different applied stresses with the same parameters.

From Fig. 2, we note that the values of Z for σ = 10MPa are higher than those for σ = 0MPa at the whole interval of frequency. For higher values of stresses (σ >10MPa), Z decreases monotonically with σ at low frequencies. At higher frequencies, the impedance curves corresponding to a particular value of stress crosses the σ = 0MPa curve thereby making Z(σ)>Z(0) implying that the SI ratio is positive within these intervals of frequency. The inset shows the complex behavior of susceptibility as functions of different frequencies and applied stresses. The real component of susceptibility ($\chi'$) initially remains constant with frequency but decreases to a large extent at higher frequencies. Further the values of $\chi'$ corresponding to 10MPa are higher than those of 0MPa for all frequencies which may be correlated with the behavior of Z. The imaginary component of susceptibility ($\chi''$) increases from very low values at low frequencies, exhibits maxima at intermediate frequencies and then decreases monotonically. With the increase of stress, the peak value of imaginary susceptibility ($\chi''_{max}$) also becomes a function of applied stress and it gets shifted to higher frequencies with the application of stress.

We now discuss the stress-impedance effect as a function of different orientations ($\theta_0$) of anisotropy field $H_a$. From equation (18) in our model which governs the equilibrium position of **M**, we see that $\theta_0$ determines the orientation of the magnetization vector. Fig.3 shows the variation of relative change in impedance as functions of applied stresses and for different values of $\theta_0$ at 100KHz frequency, where $\theta_0$ is the angle between the anisotropy field ***H**a* and the direction of stress (z-axis). From Fig.3, we find that the orientation of anisotropy has a major role to play in SI effect. The impedance response for low values of $\theta_0$ (up to $50^0$) is entirely different from those of still higher values. Up to $\theta_0 = 50^0$, Z increases at lower values of stress, exhibits a maxima and then decreases monotonically at higher stresses. The position of impedance maxima shifts to lower values of stress as $\theta_0$ is increased to $50^0$. This behavior changes entirely when $\theta_0 > 50^0$ where Z decreases monotonically with the applied stress. The sharpness of fall of Z also increases with $\theta_0$ in the regime of $\theta_0 > 50^0$ and is maximum for $\theta_0 = 90^0$. So the magnetoelasitc film would obtain the maximum SI effect if the applied stress is perpendicular to the uniaxial magnetic anisotropy. And we can find that a bias magnetic field has always been applied perpendicular to the longitudinal direction of the magnetoelastic film to improve the SI effect in the experimental reports [14-16].

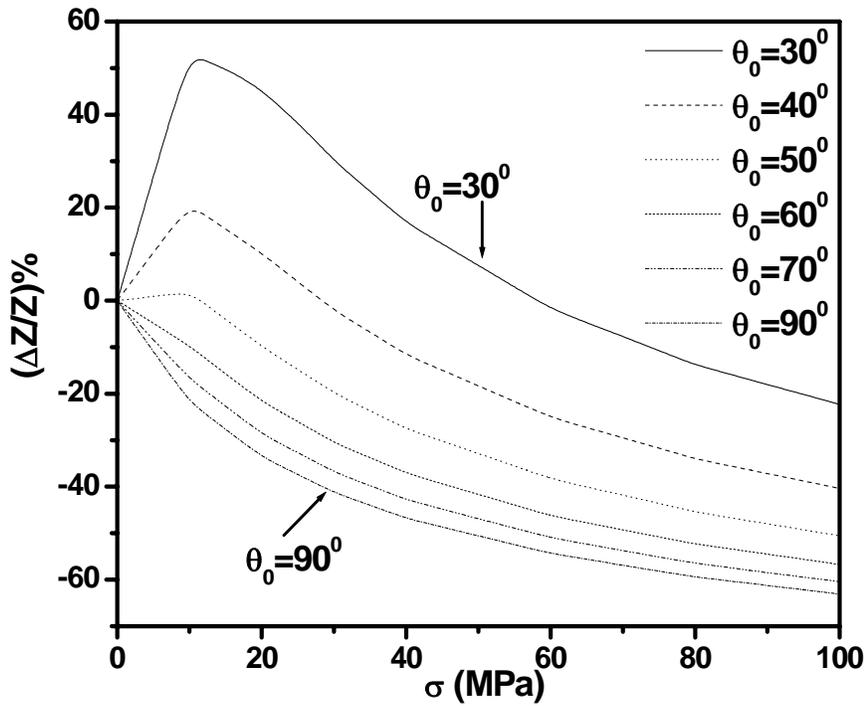

Fig. 3. Dependence of SI effect on direction of uniaxial anisotropy ($\theta_0$) at 100KHz frequency. The values of parameters taken are $\theta_0 = 50^0$, $H_a = 200$ A/m, $\alpha=20$, $M_s = 6.5 \times 10^5$ A/m and $\omega_0 = 10^6$ rad/s.

In Fig.4, we depict the variation of relative change in magneto-impedance with external biasing magnetic field, H (normalized with respect to the anisotropy field $H_a$ = 200A/m) at different applied stresses and at 1MHz frequency.

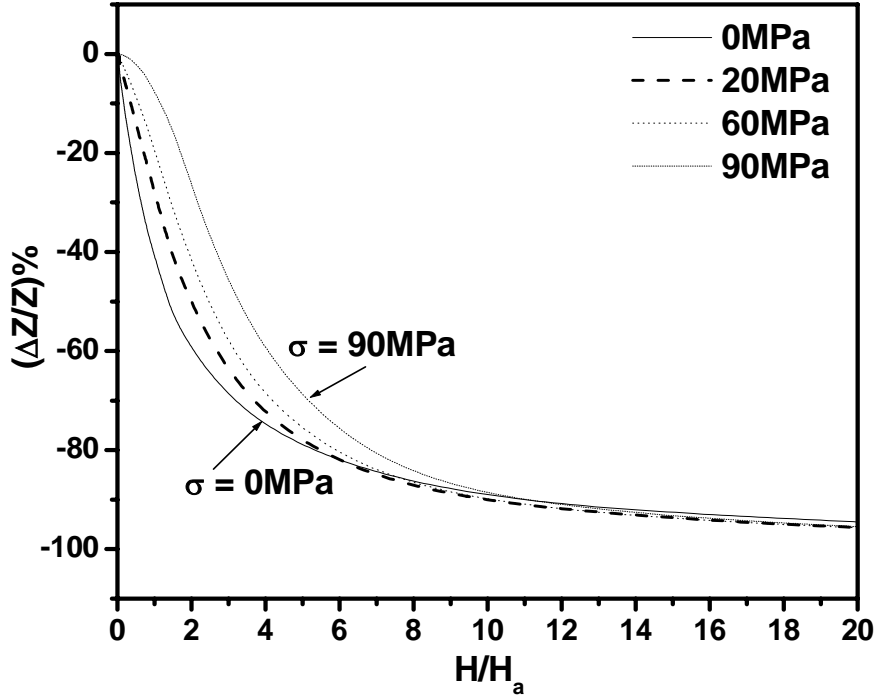

Fig.4. Variation of relative change in magneto-impedance [$(\Delta Z/Z)\%$ = 100x(Z(H)-Z(0)/Z(0))] with normalized field $H/H_a$ ($H_a$= 200A/m) at different applied stresses and at 1MHz frequency. The values of various parameters are $\theta_0 = 50^0$, $H_a$ = 200A/m, $\alpha$ = 20, $M_s$ = 6.5x10$^5$A/m, $\omega_0 = 10^6$ rad/s $\gamma_1$ = -7x10$^{-18}$A$^{-2}$m$^2$ and $\gamma_1'(0) = 1x10^{-29} A^{-2}m^2 Pa^{-1}$.

The impedance exhibits a maximum at zero bias field and decreases with the increase of bias field, thus exhibiting negative magneto-impedance. The results show a maximum MI ratio of about 99% for all stresses. With the increase of σ, the MI response becomes flatter and it saturates at higher fields. This MI behavior under the influence of stress is common to negative magnetostrictive materials, since impedance decreases with the application of stress in them [4,5].

The stress-impedance behavior, in particular the dependence of Z on applied stresses at various excitation frequencies obtained from the theoretical model is shown in Fig.5.

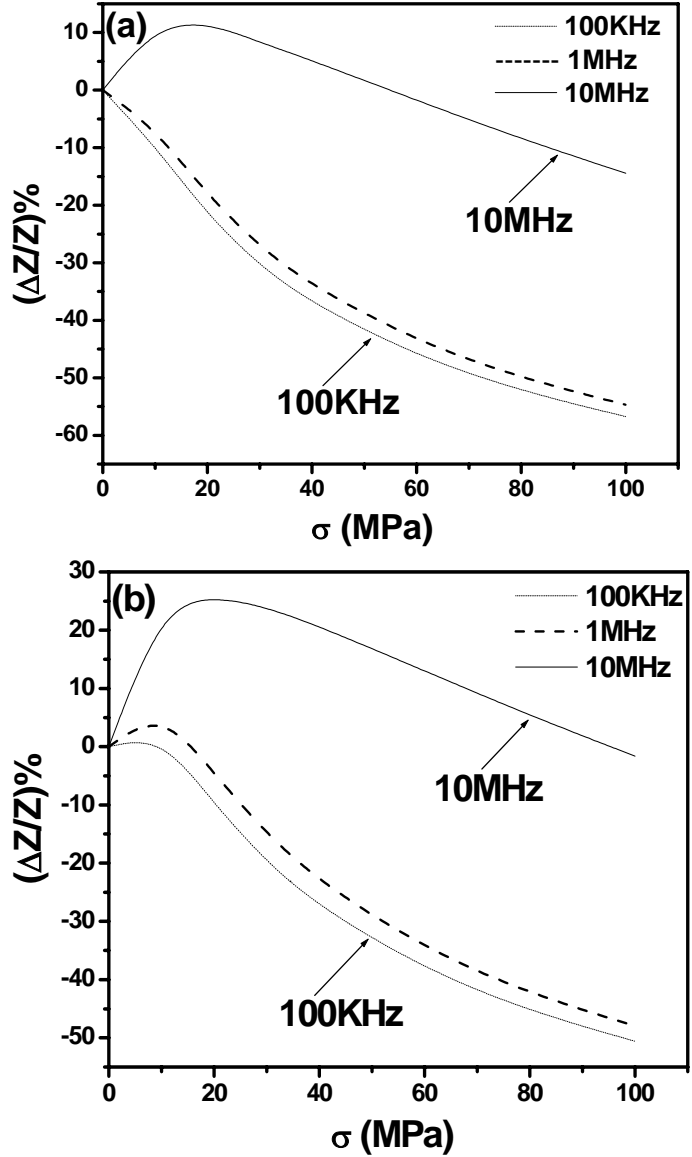

Fig.5. The stress dependence of relative change in impedance ($\Delta Z/Z$) % as functions of excitation frequencies obtained from the theoretical model with $\theta_0 = 60^0$ (a) and $\theta_0=50^0$ (b). All the other parameters $H_a = 200$ A/m, $\alpha = 20$, $M_s = 6.5 \times 10^5$ A/m, $\omega_0 = 10^6$ rad/s, $\gamma_1 = -7 \times 10^{-18} A^{-2}m^2$ and $\gamma_1^{'}(0)=1 \times 10^{-29} A^{-2}m^2 Pa^{-1}$ were kept fixed.

Fig.5 depicts the frequency dependence of SI effect at two different values of $\theta_0$. Fig. 5(a) correspond to $\theta_0 = 60^0$ and 5(b) to $\theta_0 = 50^0$. All the other parameters have been kept fixed. At low frequencies, impedance decreases monotonically with stress but at higher frequencies ($\geq 1$MHz), the SI curves exhibits a peak where the SI ratio becomes positive and then decreases at higher stresses. Such dependences of SI curves on external frequencies have also been observed experimentally in $Co_{71-x}Fe_xCr_7Si_8B_{14}$ (x = 0, 2) amorphous ribbons [4, 5].

In Fig.6, we depict the effect of change in magnitude of anisotropy field ($H_a$) in the stress-impedance behavior according to our model. The values of different parameters taken are $\theta_a = 50^0$, $H_a = 200$A/m, $\alpha=20$, $M_s = 6.5 \times 10^5$ A/m and $\omega_0 = 10^6$ rad/s.

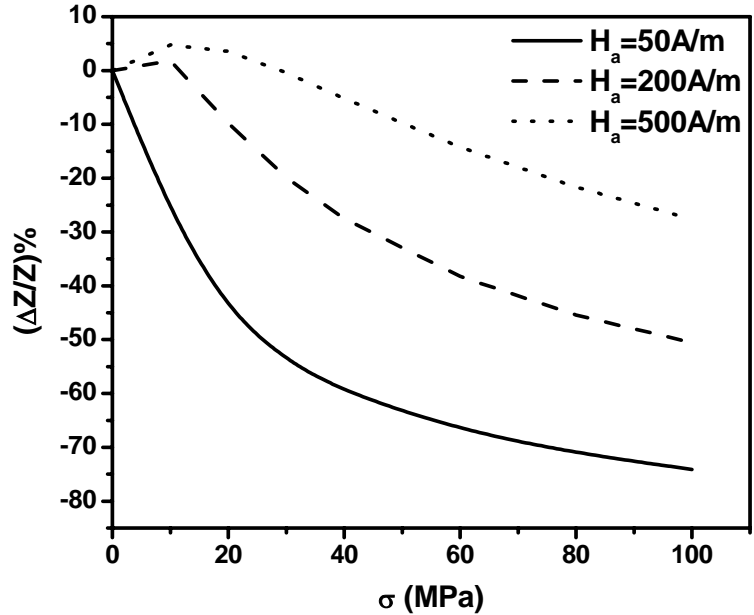

Fig.6. Dependence of SI effect upon magnitudes of anisotropy field, $H_a$ at 100KHz frequency. The values of parameters are $\theta_a = 50^0$, $H_a = 200$A/m, $\alpha=20$, $M_s = 6.5 \times 10^5$ A/m and $\omega_0 = 10^6$ rad/s.

The influence of anisotropy field in the SI behavior is clear from the plot. With the increase of $H_a$, the SI behavior becomes flatter due to the decrease in impedance values. At higher values of $H_a$, the SI ratio becomes positive at small stresses and decreases at higher stresses.

In Fig.7, we depict the influence of two damping parameters, the Gilbert damping term (denoted by α) and Bloch-Bloembergen damping term (denoted by $\tau = \frac{1}{\omega_0}$) on the SI response in accordance with our model.

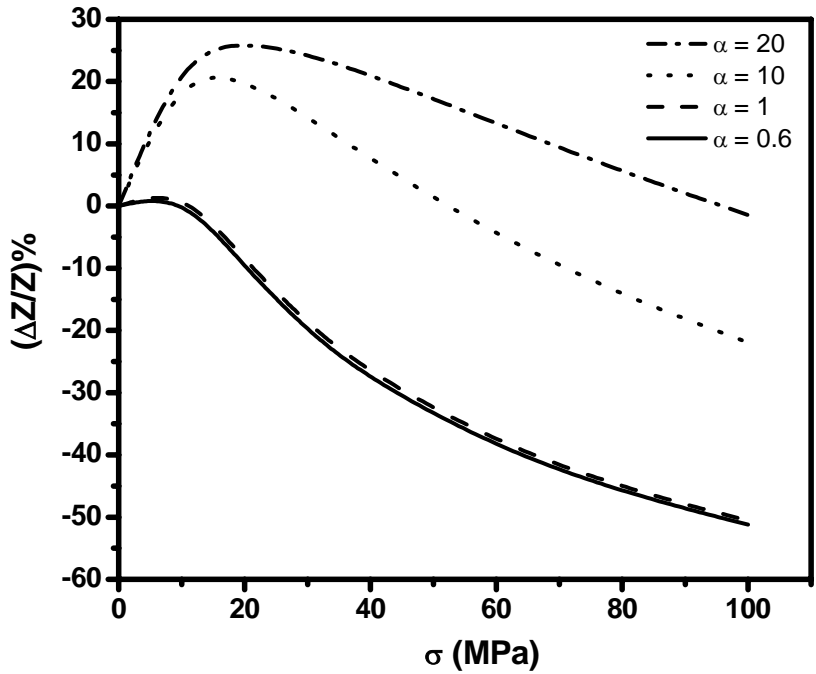

Fig.7(a). Influence of Gilbert damping term on the SI response according to the model. The values of various parameters are $\theta_0 = 50^0$, $H_a = 200$A/m, $M_s = 6.5 \times 10^5$A/m, $\omega_0 = 10^6$ rad/s, $\gamma_I(0) = -7 \times 10^{-18} A^{-2} m^2$ and $\gamma_I'(0) = 1 \times 10^{-29} A^{-2} m^2 Pa^{-1}$. The frequency is 10MHz.

Fig.7(a) shows the influence of α on the SI response according to the theoretical model. The excitation frequency was kept fixed at 10MHz. The values of various parameters are $\theta_0 = 50^0$, $H_a = 200$A/m, α = 20, $M_s = 6.5 \times 10^5$A/m, $\omega_0 = 10^6$ rad/s, $\gamma_1(0) = -7 \times 10^{-18} A^{-2} m^2$ and $\gamma_1'(0) = 1x10^{-29} A^{-2} m^2 Pa^{-1}$. The results show that with the increase of α (from 0.6 to 20), the SI response becomes flatter and the maximum relative change in SI decreases with α.

Keeping the other parameters fixed the influence of Bloch-Bloembergen damping term on the SI response is depicted in Fig.7(b).

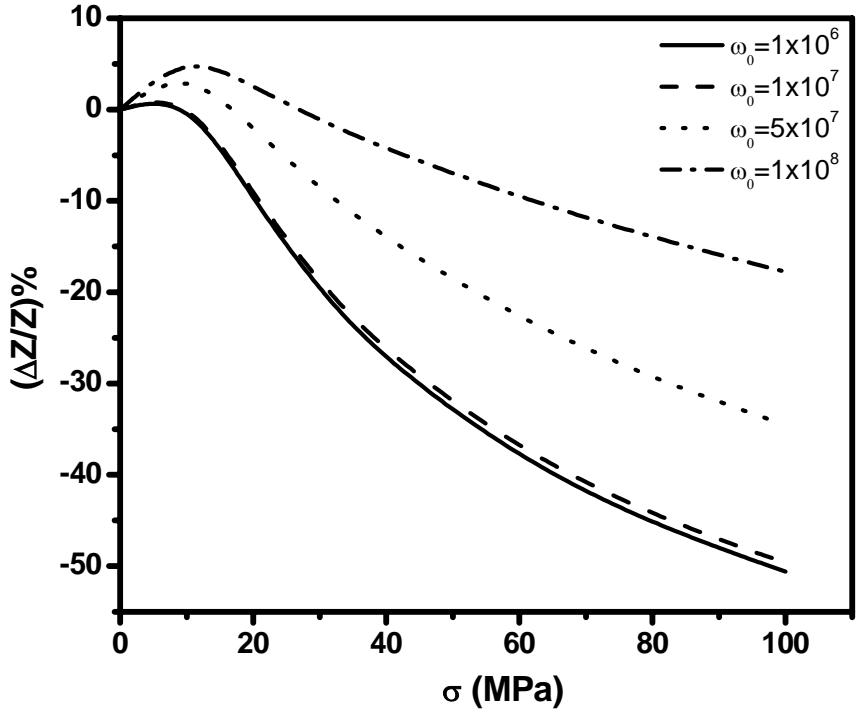

Fig.7(b). Influence of Bloch-Bloembergen damping term on the SI response according to the model. The values of various parameters are $\theta_0 = 50^0$, $H_a = 200 A/m$, α = 1, $M_s = 6.5x10^5 A/m$, $\gamma_1 = -7x10^{-18} A^{-2} m^2$ $\gamma_1'(0) = 1x10^{-29} A^{-2} m^2 Pa^{-1}$. The frequency is kept fixed at 1MHz. The units of $\omega_0$ are in rad/s.

In Fig.7(b), the Gilbert damping parameter, α, is kept at a constant value of 1 and the excitation frequency is 1MHz. The results show the broadening of SI curves (very similar to those in Fig.7(a)), thus decreasing the maximum relative change in SI with the increase of $\omega_0$ (from $10^6$ to $10^8$ rad/s).

## 4. Conclusions

In this paper, based on the Landau-Lifsitz-Gilbert equation, the stress-impedance effects in a negative magnetostrictive film as functions of several extrinsic and intrinsic parameters have been studied. The results show that SI effects are strongly dependent upon excitation frequency exhibiting a peak in high frequency regime. The orientation of anisotropy ($\theta_0$) also plays an important role in determining the nature of variation of SI. The results show that maximum SI effect can be obtained if the stress is applied perpendicular to the direction of uniaxial magnetic anisotropy.

## List of Figure Captions

Fig. 1. Schematic view of magnetoelastic film showing different directions of anisotropy fields and stress.

Fig.2. Frequency dependence of Z as functions of applied stresses. The values of parameters taken are $\theta_0 = 50^0$, $H_a = 200$A/m, $\alpha=20$, $M_s = 6.5 \times 10^5$ A/m, and $\omega_0 = 10^6$ rad/s. Inset shows the frequency response of susceptibility at different applied stresses with the same parameters.

Fig. 3. Dependence of SI effect on direction of uniaxial anisotropy ($\theta_0$) at 100KHz frequency. The values of parameters taken are $\theta_0 = 50^0$, $H_a = 200$A/m, $\alpha=20$, $M_s = 6.5 \times 10^5$ A/m and $\omega_0 = 10^6$ rad/s.

Fig.4. Variation of relative change in magneto-impedance [$(\Delta Z/Z)\% = 100 \times (Z(H)-Z(0)/Z(0))$] with normalized field $H/H_a$ ($H_a = 200$A/m) at different applied stresses and at 1MHz frequency. The values of various parameters are $\theta_0 = 50^0$, $H_a = 200$A/m, $\alpha = 20$, $M_s = 6.5 \times 10^5$A/m, $\omega_0 = 10^6$ rad/s $\gamma_1(0) = -7 \times 10^{-18} A^{-2}m^2$ and $\gamma_1'(0) = 1 \times 10^{-29} A^{-2}m^2 Pa^{-1}$.

Fig.5. The stress dependence of relative change in impedance ($\Delta Z/Z$) % as functions of excitation frequencies obtained from the theoretical model with $\theta_0 = 60^0$ (a) and $\theta_0 = 50^0$

(b). All the other parameters $H_a$ = 200A/m, α = 20, $M_s$ = 6.5x10$^5$A/m, $\omega_0$ = 10$^6$ rad/s, $\gamma_1(0) = -7 \times 10^{-18} A^{-2} m^2$ and $\gamma_1'(0) = 1 \times 10^{-29} A^{-2} m^2 Pa^{-1}$ were kept fixed.

Fig.6. Dependence of SI effect upon magnitudes of anisotropy field, $H_a$ at 100KHz frequency. The values of parameters are $\theta_a$ = 50$^0$, $H_a$ =200A/m, α=20, $M_s$ = 6.5x10$^5$ A/m and $\omega_0$ = 10$^6$ rad/s.

Fig.7(a). Influence of Gilbert damping term on the SI response according to the model. The values of various parameters are $\theta_0$ = 50$^0$, $H_a$ = 200A/m, $M_s$ = 6.5x10$^5$A/m, $\omega_0$ = 10$^6$ rad/s, $\gamma_1(0) = -7 \times 10^{-18} A^{-2} m^2$ and $\gamma_1'(0) = 1 \times 10^{-29} A^{-2} m^2 Pa^{-1}$. The frequency is 10MHz.

Fig.7(b). Influence of Bloch-Bloembergen damping term on the SI response according to the model. The values of various parameters are $\theta_0$ = 50$^0$, $H_a$ = 200A/m, α = 1, $M_s$ = 6.5x10$^5$A/m, $\gamma_1$= -7x10$^{-18}$A$^{-2}$m$^2$ $\gamma_1'(0) = 1 \times 10^{-29} A^{-2} m^2 Pa^{-1}$. The frequency is kept fixed at 1MHz. The units of $\omega_0$ are in rad/s.